\documentclass[aps,prb,10pt,amsmath,amssymb,twocolumn,floatfix,superscriptaddress,showpacs,numerical,footinbib]{revtex4-1}

\usepackage{graphicx}
\usepackage[pdftex]{color}
\usepackage[colorlinks,bookmarks=true,citecolor=blue,linkcolor=red,urlcolor=blue]{hyperref}

\newcommand{\ket}[1]{| #1 \rangle}
\newcommand{\bra}[1]{\langle #1 |}
\newcommand{\braket}[2]{\langle #1 | #2 \rangle}

\newcommand{\sgn}[1]{\mathrm{sgn} \left(#1 \right)}

\begin{document}
\title{Transversal magnetotransport in Weyl semimetals: Exact numerical approach}
\author{Jan Behrends}
\affiliation{Max-Planck-Institut f\"ur Physik komplexer Systeme, 01187 Dresden, Germany}
\author{Flore K. Kunst}
\affiliation{Department of Physics, Stockholm University, AlbaNova University Center, 106 91 Stockholm, Sweden}
\author{Bj\"orn Sbierski}
\affiliation{Dahlem Center for Complex Quantum Systems and Institut f\"ur Theoretische Physik, Freie Universit\"at Berlin, 14195 Berlin, Germany}

\begin{abstract}
Magnetotransport experiments on Weyl semimetals are essential for investigating the intriguing topological and low-energy properties of Weyl nodes.
If the transport direction is perpendicular to the applied magnetic field, experiments have shown a large positive magnetoresistance.
In this work, we present a theoretical scattering matrix approach to transversal magnetotransport in a Weyl node. Our numerical method confirms and goes beyond the existing perturbative analytical approach by treating disorder exactly.
It is formulated in real space and is applicable to mesoscopic samples as well as in the bulk limit. In particular, we study the case of clean and strongly disordered samples.    
\end{abstract}

\maketitle

\section{Introduction}

Materials realizing a nodal dispersion, such as Dirac and Weyl semimetals, are currently one of the main research topics in condensed matter physics.
Early proposals for materials featuring Weyl nodes\cite{Wan2011,Weng2015,Huang:2015ig} were soon followed by spectroscopic experiments confirming the nodal bulk dispersions and hallmark surface states, which appear in the form of Fermi arcs.\cite{Xu2015a,Lv2015a,Xu2015,Bernevig2015}
Weyl materials show peculiar magnetotransport properties\cite{Hosur2013} that are reflected in two distinct experimental observations.
First, when the electric field is aligned parallel to the magnetic field, i.e., $\mathbf{E}\parallel\mathbf{B}$, the observed unusual negative magnetoresistance\cite{Xiong:2015kl,Li:2016bj,Xiong:2016bx,Li:2016hs,Li:2016dp,Zhang:2016cv,Wang:2016hc,Wang:2017eo} is understood in terms of the chiral anomaly.\cite{Nielsen:1983ce,Son2013}
In a simplified picture, as a magnetic field turns the Weyl node into a highly degenerate chiral state, backscattering has to occur at the inter-node level.
This is much less effective than intra-node scattering relevant for the $B=0$ case and hence leads to an increased conductivity.
Second, and of main interest for the rest of the paper, we consider transversal magnetotransport, i.e., $\mathbf{E}\perp\mathbf{B}$, where the chiral anomaly is not operational.
Experimentally, a large transversal magnetoresistivity has been observed in a variety of Dirac and Weyl materials at low temperatures around $T=2\,\mathrm{K}$.
Early studies of the Dirac semimetal Cd$_{3}$As$_{2}$\cite{Feng2015,Liang2015,Zhao2015,Li:2016bj} with multiple Dirac cones were followed by measurements on the single-Dirac cone material TiBiSSe \cite{Novak2015} and Weyl materials NbAs\cite{Ghimire2015}, TaAs\cite{Huang2015b} and NbP.\cite{Shekhar2015}
Note that a Dirac node splits into two Weyl nodes when a magnetic field is applied.
The transversal resistivity can increase up to a factor of a thousand compared to the $(B=0)$-resistivity upon the application of a magnetic field with a strength on the order of $10\,\mathrm{T}$.\cite{Novak2015}

\begin{figure}[b]
\includegraphics[width=\linewidth]{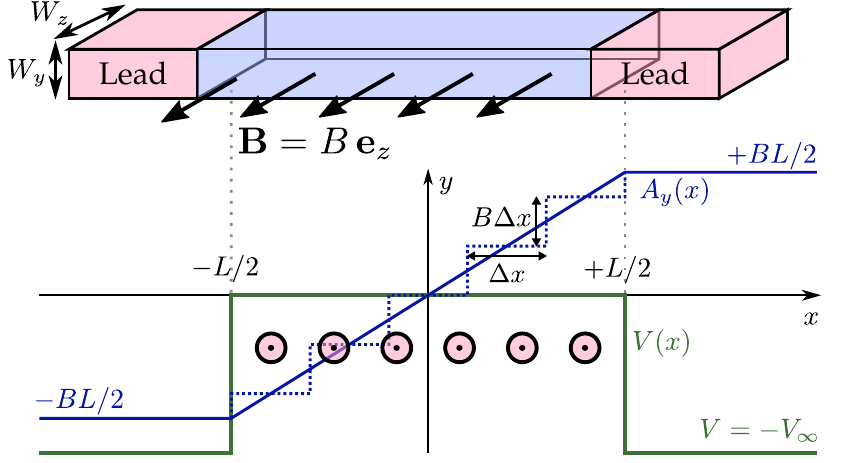}
\caption{\label{fig:setup} Sketch of modeled transport experiment.
The magnetic field points in the $z$-direction perpendicular to the transport direction ($x$).
The lower panel shows the doping distribution $V(x)$ (green) that defines the leads, and the choice for the vector potential $A_{y}$ (blue).
The piecewise constant approximation of $A_{y}$, shown with a dotted dotted line, is used in the scattering matrix approach for the situation including disorder $U(\mathbf{r})$.
}
\end{figure}

The observed linear and unsaturated growth of the resistivity with the applied magnetic field is in stark contrast to Boltzmann magnetotransport theory in metals where the magnetoresistivity is much smaller, quadratic in $B$ for small $B$ and saturating if the cyclotron frequency exceeds the scattering time.
Thus, alternative theoretical approaches are necessary to explain the experimental results. 

Abrikosov studied the problem of transverse magnetoresistance long before the confirmation of Weyl and Dirac materials in the laboratory.
He developed a perturbative analytical approach for transversal magnetotransport of a single Weyl node Hamiltonian.\cite{Abrikosov1998}
He used the Kubo formula and Born approximation to include the effect of weak disorder while focusing on the quantum limit.
Indeed, this approach results in a linear and non-saturating magnetoresistance under the assumption of screened Coulomb impurities.

The recent experiments triggered a renewed interest in the problem
leading to the appearance of a number
of theoretical studies reproducing and extending Abrikosov's earlier results.\cite{Gorbar2014,Lu2015c,Klier2015,Zhang2016,Xiao2016,Klier2017,Ji2017,Cortijo:2016bq} Notably,
the more realistic situation with a pair of nodes was accounted
for in Refs.~\onlinecite{Gorbar2014,Lu2015c,Zhang2016,Ji2017}, while disorder was
assumed to be of (uncorrelated) Gaussian white noise type
in Refs.~\onlinecite{Lu2015c,Klier2015,Klier2017} and to have a finite range in Ref.~\onlinecite{Zhang2016}. However,
most of the approaches in the recent papers follow Abrikosov's method by treating disorder using the Born approximation and computing the conductivity making use of the Kubo formula with bare current vertices. Some slight but noteworthy variations are the numerical evaluation of overlap integrals for disorder scattering in Ref.~\onlinecite{Xiao2016} and the extension to the self-consistent Born approximation (SCBA) in Refs.~\onlinecite{Klier2015,Klier2017}. Moreover, the role of vertex corrections was scrutinized in Refs.~\onlinecite{Klier2015,Klier2017}. 
Staying in the quantum limit, all of these recent works have confirmed Abrikosov's result for the transversal magnetoconductivity, $\sigma_{xx}(B)\sim1/B$ (screened Coulomb disorder) and $\sigma_{xx}(B)\sim B$ (short-range correlated disorder). 
Results for finite chemical
potential $\mu$ and finite temperature $T$ are also available.\cite{Xiao2016,Klier2017}
However, the extent to which the above theories are applicable to experiments is still under debate.\cite{Xiao2016}

In this work, we complement the existing analytical and perturbative approaches to transversal magnetotransport in a Weyl node by introducing a numerical method that treats disorder and the magnetic field exactly.
This is of relevance in the limit $B \rightarrow 0$ due to the known failure of the SCBA, which misses relevant crossed disorder diagrams.\cite{Sbierski2014a}
It also allows us to study the interplay between magnetic field and strong disorder, the latter of which is known to drive a phase transition\cite{Fradkin1986,Syzranov2016c} between a semimetal and a diffusive metal in the case $B=0$.
Finally, by being formulated in the framework of scattering theory, our formalism applies to mesoscopic systems while the bulk limit is well within reach.

Although we formulate our method for undoped Weyl nodes $\mu=0$, it can be straightforwardly generalized to include finite $\mu$. Similarly, although the disorder type could be freely specified, we limit ourselves to Gaussian disorder with correlation length $\xi$.

After introducing the model in Sec.~\ref{sec:model}, we apply our method in three different situations:
first, in Sec.~\ref{sec:clean}, we extensively discuss the limit of mesoscopic magnetotransport in a clean Weyl node.
Second, we introduce the exact treatment of disorder in Sec.~\ref{sec:numerical} and review the perturbative Born-Kubo approach in Sec.~\ref{sec:Born-Kubo}.
We show in Sec.~\ref{sec:results} that for weak disorder both theories are in excellent agreement.
For strong disorder we find a negligible dependence of the conductivity on the magnetic field.
We conclude in Sec.~\ref{sec:conclusion}.

\section{Model \label{sec:model}}

We study magnetotransport in a Weyl-semimetal slab with length $L$ and transversal widths $W_y, W_z \gg L$ shown schematically in Fig.~\ref{fig:setup} (drawing not to scale).
The scattering between several Weyl nodes can be neglected if their separation in reciprocal space is large compared to the inverse disorder correlation length and we consequently focus on a single node.
We assume the magnetic field $\mathbf{B}=B\mathbf{e}_{z}$
in the $z$-direction
and transport in the $x$-direction.
The Hamiltonian reads
\begin{equation}
H=v\left(\mathbf{p}-e\mathbf{A}\right)\cdot\boldsymbol{\sigma}+V\left(x\right)+U\left(\mathbf{r}\right),
\end{equation}
where $v$ is the Fermi velocity, $\mathbf{p}$ the momentum operator
and the vector potential is written in the Landau gauge
\begin{equation}
\mathbf{A}=\begin{cases}
-B\frac{L}{2}\,\mathbf{e}_{y} & :\,x\leq-\frac{L}{2},\\
Bx\,\mathbf{e}_{y} & :\,|x|<\frac{L}{2},\\
B\frac{L}{2}\,\mathbf{e}_{y} & :\,x\geq\frac{L}{2}.
\end{cases}
\label{eq:exactA-1}
\end{equation}
The leads at $|x|\geq L/2$ are assumed to be free of magnetic field and are modeled as highly doped Weyl metals with the potential
$V(x)$,
\begin{equation}
V(x)=\begin{cases}
0 & :\,|x|<L/2,\\
-V_{\infty} & :\,\mathrm{otherwise}.
\end{cases}
\end{equation}
In the central scattering region, the Fermi energy is located at the nodal point.
The disorder potential $U\left(\mathbf{r}\right)$ is assumed to be present in the scattering region only and is modeled as a Gaussian correlated with zero mean and
\begin{equation}
 \left\langle U(\mathbf{r})U(\mathbf{r}^{\prime})\right\rangle _\mathrm{dis}=\frac{K\left(\hbar v\right)^{2}}{(2\pi)^{3/2}\xi^{2}}\exp\left(-\frac{|\mathbf{r}-\mathbf{r}^{\prime}|^{2}}{2\xi^{2}}\right)
\label{eq:disorderCorrelator}
\end{equation}
where $\left\langle \cdots \right\rangle_\mathrm{dis}$ denotes a disorder average and $\xi$ is the disorder correlation length.
The disorder strength is measured by the dimensionless parameter $K$.

\section{Mesoscopic transport in clean samples\label{sec:clean}}

We start by considering transport in the clean limit, i.e., $K=0$.
In the transversal $y$ and $z$-directions, we apply periodic boundary conditions and use the resulting translational symmetry to make the ansatz $\psi(\mathbf{r})=\psi(x)e^{i\left(k_{y}y+k_{z}z\right)}$ with $\psi(x)=\left(\psi_{\uparrow}(x),\psi_{\downarrow}(x)\right)^{\mathrm{T}}$.
We solve the scattering problem by assuming an incoming state from $x=-\infty$ and finding the transmission coefficient $t$.
In the central scattering region, $|x|\leq L/2$, the Schr\"{o}dinger equation for a zero-energy state leads to the following system of equations
\begin{align}
\partial_{x}\psi_{\uparrow}(x) & =+\left(k_{y}-xl_{B}^{-2}\right)\psi_{\uparrow}(x)+ik_{z}\psi_{\downarrow}(x),\nonumber \\	
\partial_{x}\psi_{\downarrow}(x) & =-\left(k_{y}-xl_{B}^{-2}\right)\psi_{\downarrow}(x)-ik_{z}\psi_{\uparrow}(x),
\end{align}
where $l_{B}=\sqrt{\hbar/(eB)}$ is the magnetic length.
In the limit of infinite doping in the leads, $V_{\infty}\rightarrow\infty$, the following boundary conditions are enforced by the lead states from wave-function matching
\begin{equation}
\psi(-L/2)=\frac{1}{\sqrt{2}}\left(\begin{array}{c}
1\\
1
\end{array}\right)+\frac{r}{\sqrt{2}}\left(\begin{array}{c}
1\\
-1
\end{array}\right),\label{eq:BC_L-1}
\end{equation}
with $r$ the reflection coefficient and 
\begin{equation}
\psi(L/2)=\frac{t}{\sqrt{2}}\left(\begin{array}{c}
1\\
1
\end{array}\right).\label{eq:BC_R-1}
\end{equation}
The two spinors belong to the left- and right-propagating modes in the leads, i.e., the eigenvectors of the velocity operator $v \sigma_x$.
To obtain solutions for $t$, we redefine $\psi_{\pm}(x)\equiv\psi_{\uparrow}(x)\pm\psi_{\downarrow}(x)$,
such that the coupled system of equations reads
\begin{align}
\partial_{x}\psi_{\pm}(x) & =\left(k_{y}-xl_{B}^{-2}\mp ik_{z}\right)\psi_{\mp}(x),\label{eq:D_psiPm}
\end{align}
and the boundary conditions become $\psi_{-}(L/2)=0$ and $\psi_{+}(-L/2)=2/\sqrt{2}$. The transmission coefficient $t=\psi_{+}(L/2)/\sqrt{2}$ can be found by numerically solving Eq.~(\ref{eq:D_psiPm}).

\begin{figure}
\includegraphics[width=\linewidth]{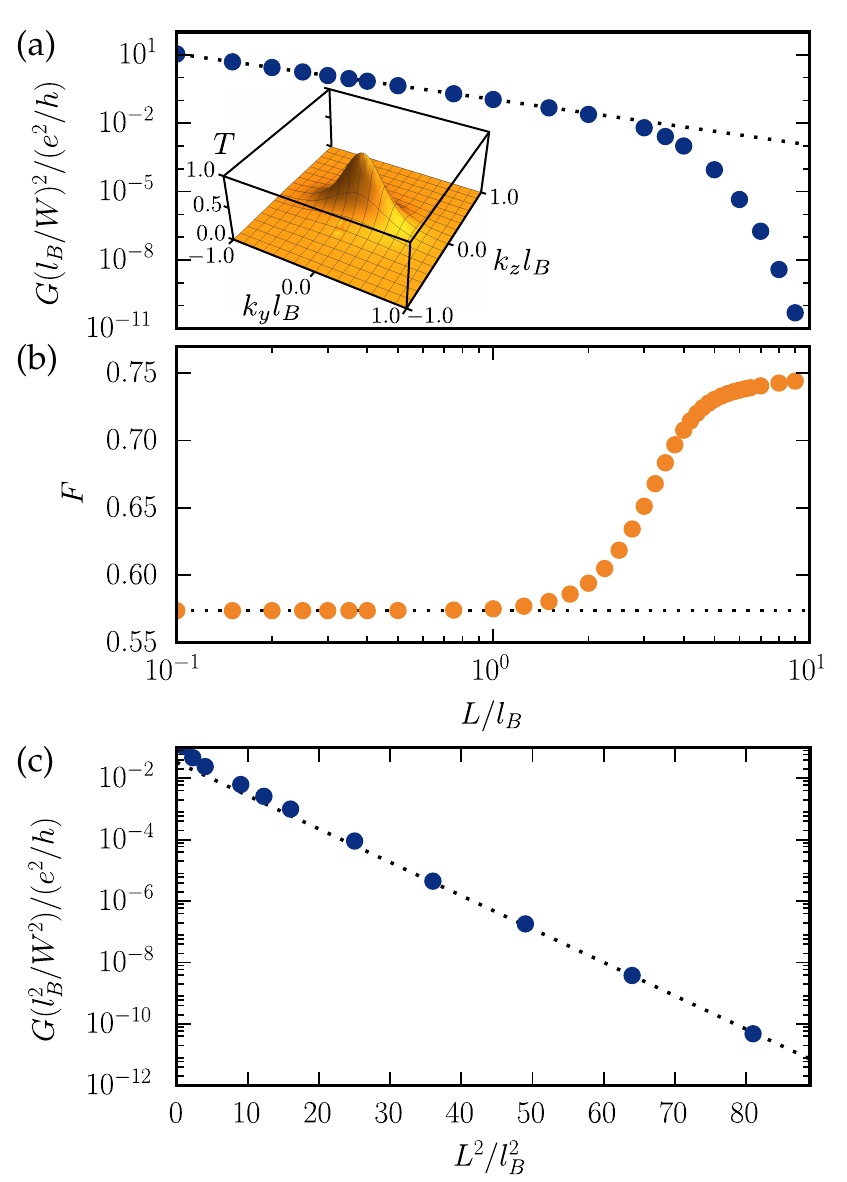}
\caption{\label{fig:clean_transport} Numerical results for clean transversal
magnetotransport.
(a) Conductance versus system length, the dotted line indicates the $B=0$ result $G\sim L^{-2}$.
The inset shows the transmission eigenvalue $T$ as a function of transversal momenta for $L/l_{B}=3$.
(b) The Fano factor versus system length. The dotted line corresponds to the $B=0$ result, $F = 0.57$.
(c) Conductance  versus $(L/l_{B})^{2}$. For the case $L/l_{B}\gg 1$, the logarithm of the conductance obeys a linear relation with $(L/l_{B})^{2}$.
The slope of the dotted line is $-1/4$.}
\end{figure}

In the inset of Fig.~\ref{fig:clean_transport}(a), the transmission eigenvalue
$T=|t|^{2}$ is shown as a function of $(k_{y},k_{z})\equiv \mathbf{k}_\perp$ for $L/l_{B}=3$.
We observe that the presence of the magnetic field causes a gauge-dependent $k_{y}$-$k_{z}$ asymmetry whereas the $B=0$ result $T=\cosh^{-2}\left(L|\mathbf{k_\perp}|\right)$\cite{Sbierski2014a} is rotationally symmetric. Nevertheless, the $\mathbf{k_\perp}=0$ mode still shows perfect transmission. The conductance $G$ is calculated via the Landauer formula
\begin{equation}
G = \frac{e^2}{h} {\rm tr}\left[ t t^\dagger\right],
\label{eq:G}
\end{equation}
and is shown in Fig.~\ref{fig:clean_transport}(a) where
the dotted line represents the known result in the limit $B\rightarrow0$, $G\frac{l_{B}^{2}}{W^{2}}=\frac{e^2}{h}\frac{\mathrm{ln}2}{2\pi}\left(\frac{l_{B}}{L}\right)^{2}$\cite{Sbierski2014a} with the system's transversal width $W=W_y=W_z$.
For $L/l_{B}\gg 1$, when the magnetic field becomes relevant,
the conductance vanishes rapidly with $L/l_{B}$. The plot of $\mathrm{ln}[G \left(l_{B}/W\right)^{2} h/e^2]$
vs.\ $L/l_{B}$ in Fig.~\ref{fig:clean_transport}(c) reveals a functional
form $G\frac{l_{B}^{2}}{W^{2}}\sim e^{-\alpha(L/l_{B})^{2}}$ with
$\alpha=1/4$ (dotted line). 

Although we did not succeed in deriving an analytical expression for $G$ in the limit $L/l_{B}\gg 1$, the exponential suppression can be understood from the form of the bulk wave functions in the vicinity of zero energy.
The zeroth Landau level is chiral and reads
\begin{equation}
 \psi_{0,+}\left(\mathbf{r}\right) \propto e^{ik_{y}y+ik_{z}z} \exp\left(-\frac{1}{2}\left[x/l_{B}-k_{y}l_{B}\right]^{2}\right)\left|\uparrow\right\rangle
\end{equation}
with energy $\varepsilon_{0,+}=+\hbar v k_{z}$ and $\left|\uparrow\right\rangle = (1,0)^{\mathrm{T}}$.
The $x$ position of the exponentially localized wave function is determined
by the transversal momentum $k_{y}$. Due to translational invariance
(i.e., the absence of disorder), $k_{y}$ is conserved and coherent
transport between the leads only proceeds via the exponential tail
of the orbitals leaking across the sample. This qualitatively explains the observed exponential suppression of the conductance with length.

The Fano factor
\begin{equation}
F = \frac{{\rm tr}\left[t t^\dagger \left(1 - t t^\dagger\right)\right]}{{\rm tr}\left[t t^\dagger\right]}
\label{eq:F}
\end{equation}
is a measure of shot noise in quantum transport\cite{Datta1997} and is shown in Fig.~\ref{fig:clean_transport}(b). In the limit $L \ll l_B$, we find $F = 0.57$ in agreement with the pseudo-ballistic regime described in Ref.~\onlinecite{Sbierski2014a}. With increasing field $B$, the Fano factor also increases and reaches up to $F = 0.74$ in the large-$B$ regime.

Upon the inclusion of disorder, i.e., $K>0$, the momenta are no longer conserved and we expect a diffusion process between localized orbitals leading to drastic change of transport behavior.
Indeed, we show in the next section that in the large system limit the transport with $K>0$ becomes diffusive.

\section{Numerical Magnetotransport in the presence of disorder \label{sec:numerical}}

We use a stepwise strategy to find the scattering matrix of the magnetotransport problem in the presence of a specific disorder realization.
We divide the scattering region of length $L$ in slices of length $\Delta x$, as shown in Fig.~\ref{fig:setup}, with $\Delta x$ much smaller than both $\xi$ and $l_B$.
We concentrate the magnetic flux between $x$ and $x+\Delta x$ in an infinitely thin sheet at $x+\Delta x$ where the vector potential $A_y(x)$ accordingly has a jump.
Likewise, the disorder potential in the slice is represented by an infinitely thin sheet at $x+\Delta x$. The scattering matrix $S_{x,x+\Delta x}$ of a slice can be found from concatenation of the scattering matrix $S_{x,x+\Delta x}^{(0)}$ for free propagation from $x$ to $x+\Delta x$ and the scattering matrices resulting from the vector potential step $S_{x+\Delta x}^{(B)}$ and the disorder sheet $S_{x+\Delta x}^{(\mathrm{dis})}$. The scattering matrix of the full system is found from concatenation of slice scattering matrices.
This approach to transport in the presence of disorder has been applied previously for two-dimensional Dirac dispersions\cite{Bardarson2007} and Weyl nodes,\cite{Sbierski2014a} and also in the presence of parallel magnetic fields.\cite{Behrends2017}
We refer to these references for the specific form of the disorder-sheet scattering matrix $S_{x+\Delta x}^{(\mathrm{dis})}$ and the derivation of the scattering matrix for free propagation within
the (clean and field-free) slice
\begin{equation}
S_{x,x+\Delta x}^{(0)}=\left(\begin{array}{cc}
t^{(0)} & r^{(0)\prime}\\
r^{(0)} & t^{(0)\prime}
\end{array}\right)
\end{equation}
with the following transmission and reflection blocks,
\begin{align}
t^{(0)}&=t^{(0)\prime}=\cosh[\Delta x\sqrt{\tilde{k}_{y}^{2}+k_{z}^{2}}]^{-1},\\
r^{(0)}&=\frac{\tilde{k}_{y}+ik_{z}}{\sqrt{\tilde{k}_{y}^{2}+k_{z}^{2}}}\tanh\left[\Delta x\sqrt{\tilde{k}_{y}^{2}+k_{z}^{2}}\right]=-r^{(0)\prime\star},
\end{align}
where $\tilde{k}_{y}=k_{y}-\frac{e}{\hbar}A_y(x)$ is $x$-dependent.
Finally, we compute the scattering matrix $S_{x+\Delta x}^{(B)}$ for an abrupt jump in vector
potential representing a magnetic field $B(x)=\delta(x+\Delta x) B \Delta x$.
This scattering matrix is found by considering two leads meeting at
$\tilde{x}=0$ in the presence of a jump in the vector potential,
\begin{equation}
A_{y}(\tilde{x})=\begin{cases}
A_{y,L} & :\tilde{x}<0,\\
A_{y,R} & :\tilde{x}>0.
\end{cases}
\end{equation}
At $\tilde{x}=0$, we match the wave functions (with incoming state
from $\tilde{x}=-\infty$) and use $a_{L,R}\equiv\frac{e}{\hbar}A_{yL,yR}$
and $v_{\infty}\equiv V_{\infty}/\hbar v$ to write
\begin{align}
&\mathcal{N}_{L}\begin{pmatrix}v_{\infty}+k_{z}\\
k_{x}^{L}+i(k_{y}-a_{L})
\end{pmatrix}+r\mathcal{N}_{L}\begin{pmatrix}v_{\infty}+k_{z}\\
-k_{x}^{L}+i(k_{y}-a_{L})
\end{pmatrix} \nonumber \\
& =t\mathcal{N}_{R}\begin{pmatrix}v_{\infty}+k_{z}\\
k_{x}^{R}+i(k_{y}-a_{R})
\end{pmatrix},\label{eq:A-step}
\end{align}
where $\mathcal{N}_{L,R}$ represents spinor normalizations and $k_{x}^{L,R}=\sqrt{v_{\infty}^{2}-\left(k_{y}-a_{L,R}\right)^{2}-k_{z}^{2}}$
are the momenta in transport direction in the left and right lead,
respectively. Solving Eq.~(\ref{eq:A-step}) and taking the limit
$V_{\infty}\rightarrow\infty$ as above, we find $t=1$ and $r=0$.
Thus, the scattering across a localized transverse field is trivial.
Note, however, that the magnetic field also enters the free slice scattering
matrix $S_{x,x+\Delta x}^{(0)}$ via the appearance of $A_y(x)$.
In summary, we obtain the total scattering matrix of a slice as $S_{x,x+\Delta x}^{(0)}\otimes S_{x+\Delta x}^{(\mathrm{dis})}$
(here $\otimes$ denotes scattering matrix concatenation \cite{Datta1997}).

We discretize the transversal momenta in accordance with the periodic boundary conditions, $k_{y,z}=\frac{2\pi}{W_{y,z}}m_{y,z}$, $m_{y,z}\in \mathbb{Z}$ and choose a cutoff $R_{y,z}$ such that the transversal momenta $|k_{y,z}|\xi\leq R_{y,z}$.
We are interested in the physical limit $R_{y,z}\rightarrow \infty$ and claim convergence if the conductance and Fano factor (computed from Eqs.~(\ref{eq:G}) and (\ref{eq:F}), respectively) do not vary with increased $R_{y,z}$
or reduced $\Delta x$.
The latter condition ensures that the magnetic field and disorder are taken into account exactly.
We increase transversal dimensions until $G/(W_y W_z)$ and $F$ are independent of the widths. We further check that the results do not change when antiperiodic transversal boundary conditions are applied.

We have checked that the stepwise approach presented in this section reproduces the results of Sec.~\ref{sec:clean} obtained for a smooth vector potential.
Building up the scattering matrix of the full system $S_{-L/2,L/2}$ from concatenating slices and labeling the intermediate scattering matrices $S_{-L/2,x}$ with $-L/2 \leq x \leq L/2$, we observe that the position of the maximum of the transmission $T$, denoted $(\tilde{k}_y,0)$ is at $\tilde{k}_y=-L l_B^{-2}/2$ for $x=-L/2$ and shifts to $\tilde{k}_y=0$ for $x=L/2$, cf.\ Fig.~\ref{fig:clean_transport}(a), inset.
Qualitatively, such a shift is also observed for the disordered case.
With increasing $x$, $\tilde{k}_y$ moves to larger values.
We can keep $\tilde{k}_y$ in the center of the $k_y$ mode range considered, if we apply redefinitions $S\left(k_{y},k_{z};k_{y}^{\prime},k_{z}^{\prime}\right) \to S \left(k_{y} - \delta_{y},k_{z}; k_{y}^{\prime}- \delta_{y}, k_{z}^{\prime}\right)$ along with $A(x)\to A(x)-\hbar\delta_{y}/e$ whenever $\tilde{k}_y$ increases above a threshold value.
Here, $\delta_y=2\pi/W_y$ is the mode separation.
This shift operation can be expressed as a concatenation with the scattering matrix,\cite{BardarsonMong}
\begin{equation}
S_{s}= \begin{pmatrix}
t_{s} & r_{s}^{\prime}\\
r_{s} & t_{s}^{\prime}
\end{pmatrix},
\end{equation} where
\begin{align*}
t_{s}(k_{y},k_{z};k_{y}^{\prime},k_{z}^{\prime}) & = \delta_{k_{z},k_{z}^{\prime}}\delta_{k_{y},k_{y}^{\prime}-\delta_{y}},\\
t_{s}^{\prime}(k_{y},k_{z};k_{y}^{\prime},k_{z}^{\prime}) & = \delta_{k_{z},k_{z}^{\prime}}\delta_{k_{y},k_{y}^{\prime}+\delta_{y}},\\
r_{s}(k_{y},k_{z};k_{y}^{\prime},k_{z}^{\prime}) & = e^{i\phi}\delta_{k_{z},k_{z}^{\prime}}\delta_{k_{y},k_{y,\mathrm{max}}}\delta_{k_{y}^{\prime},k_{y,\mathrm{max}}},\\
r_{s}^{\prime}(k_{y},k_{z};k_{y}^{\prime},k_{z}^{\prime}) & = e^{i\phi}\delta_{k_{z},k_{z}^{\prime}}\delta_{k_{y},k_{y,\mathrm{min}}}\delta_{k_{y}^{\prime},k_{y,\mathrm{min}}}.
\end{align*}
Here, $\phi$ is an arbitrary phase and $k_{y,\mathrm{min}}\simeq -R_y/\xi$ and $k_{y,\mathrm{max}}\simeq R_y/\xi$ are the minimal and maximal wave vectors considered.

\section{Born-Kubo analytical bulk conductivity \label{sec:Born-Kubo}}

We apply our numerical approach in the important and experimentally relevant bulk limit $L\to \infty$.
As mentioned in the introduction, under the additional assumption of weak disorder, the transversal magnetotransport is expected to be diffusive.
The conductivity $\sigma_{xx}$ can be calculated using the Kubo formula along with the Born approximation, following Abrikosov's seminal work in Ref.~\onlinecite{Abrikosov1998}. Here, weak disorder is understood to fulfill two conditions: (i) $K<K_c$ where $K_c$ is the critical disorder strength that, for $B=0$, drives the semimetal to a diffusive metal phase. From Ref.~\onlinecite{Sbierski2014a}, we know that $K_c\simeq 5$ for the specific disorder model in Eq.~(\ref{eq:disorderCorrelator}). (ii) The disorder-induced level broadening $\Gamma$ should be small compared to the Landau level separation $\sim \hbar v/l_B$. 

In the appendix, we present a self-contained derivation for the transversal magnetoconductivity in the weak disorder case for the model defined in Sec.~\ref{sec:model}. The calculation is
carried out in the limit $T\rightarrow0$ such that it can be compared to the exact numerical data but the result remains valid for finite $k_{B}T$
as long as $k_{B}T\ll\hbar v/l_B$. We find
\begin{equation}
\sigma_{xx}=\frac{e^{2}}{h\xi} \, \frac{K}{8\pi^{2}} \, \frac{1}{1+(l_{B}/\xi)^{2}} .\label{eq:sigma_xx}
\end{equation}
The disorder broadening of the lowest non-chiral Landau level is found to be 
\begin{equation}
\Gamma=\frac{K}{4\pi} \frac{\hbar v\xi}{l_{B}^{2}+\xi^{2}}.
\label{Gamma}
\end{equation}
For chemical potential at the nodal point, $\mu=0$, the Hall conductivity $\sigma_{xy}$ vanishes \cite{Abrikosov1998,Klier2015} and $\rho_{xx}=1/\sigma_{xx}.$ 

\section{Numerical results in disordered samples \label{sec:results}}

\begin{figure}
\includegraphics[width=\linewidth]{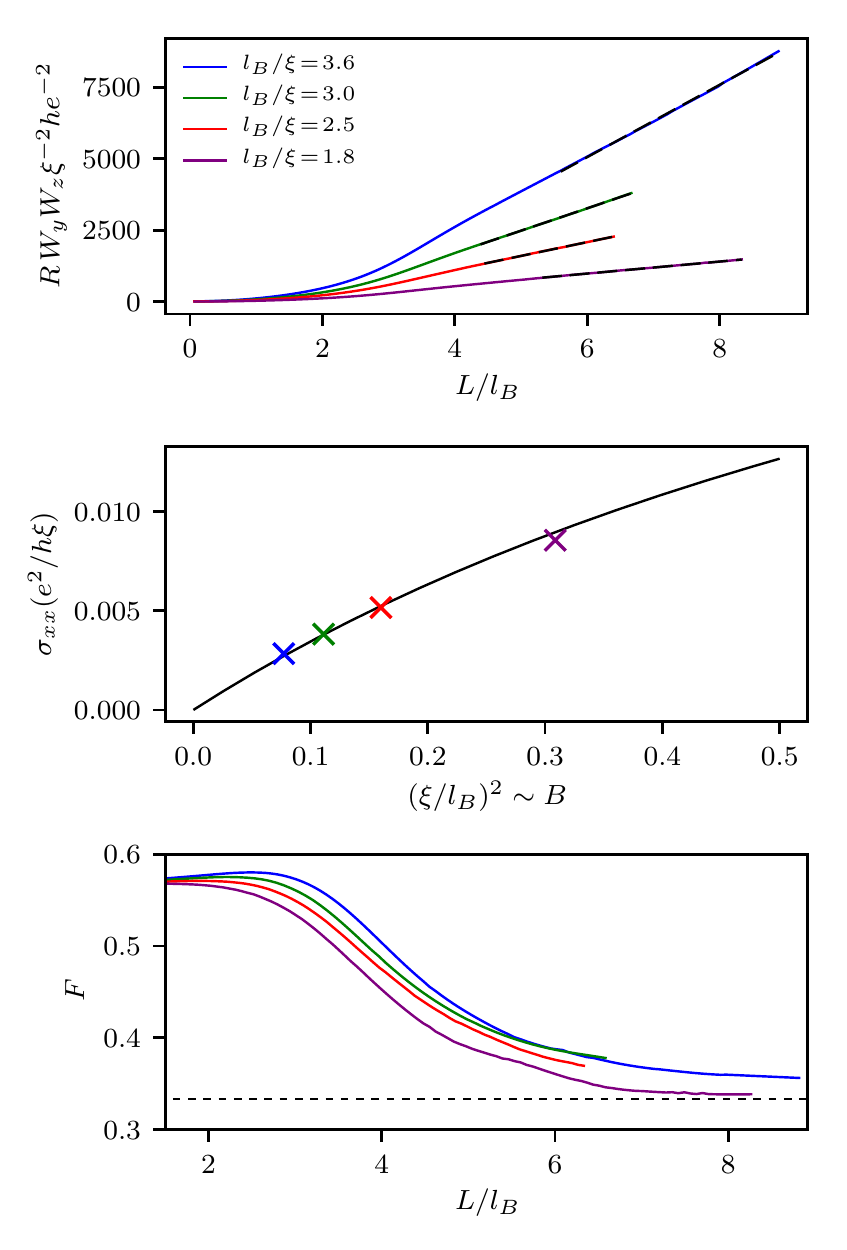}
\caption{\label{fig:weakDisorder} 
Numerical results for transversal magnetotransport in a weakly disordered Weyl node. We use $K=3$ and $l_B/\xi=1.8, 2.5, 3.0$ and $3.6$. The top panel shows the resistance averaged over $\sim\!50$ disorder realizations vs.\ length. The conductivity is extracted by a linear fit for large $L$ (dashed lines) and depicted in the middle panel. The agreement with the analytical prediction in Eq.~(\ref{eq:sigma_xx}) (solid line) is excellent. The bottom panel shows the Fano factor vs.\ length, which approaches the diffusive value of $F=1/3$ (dashed line).
}
\end{figure}

We now turn to the discussion of the results of our numerical approach from Sec.~\ref{sec:numerical} with finite disorder strength. We start with the weak disorder case and check for the validity of the analytical approach in Sec.~\ref{sec:Born-Kubo}. The numerical results for $K=3$ and $l_B/\xi=1.8, 2.5, 3.0$ and $3.6$ are presented in Fig.~\ref{fig:weakDisorder}. The top panel shows the resistance normalized to the cross-sectional area as a function of system length $L$. We find that in the bulk limit $L\rightarrow\infty$ the disorder averaged conductance behaves diffusively, i.e., $dR/dL = \mathrm{const.}$ The conductivity is extracted and depicted in the middle panel. The agreement with the analytical prediction in Eq.~(\ref{eq:sigma_xx}) (solid line) is excellent and confirms the validity of the Born-Kubo calculation. As a further confirmation of diffusive transport, the Fano factor in the bottom panel asymptotically converges to $F=1/3$.

\begin{figure}
\includegraphics[width=\linewidth]{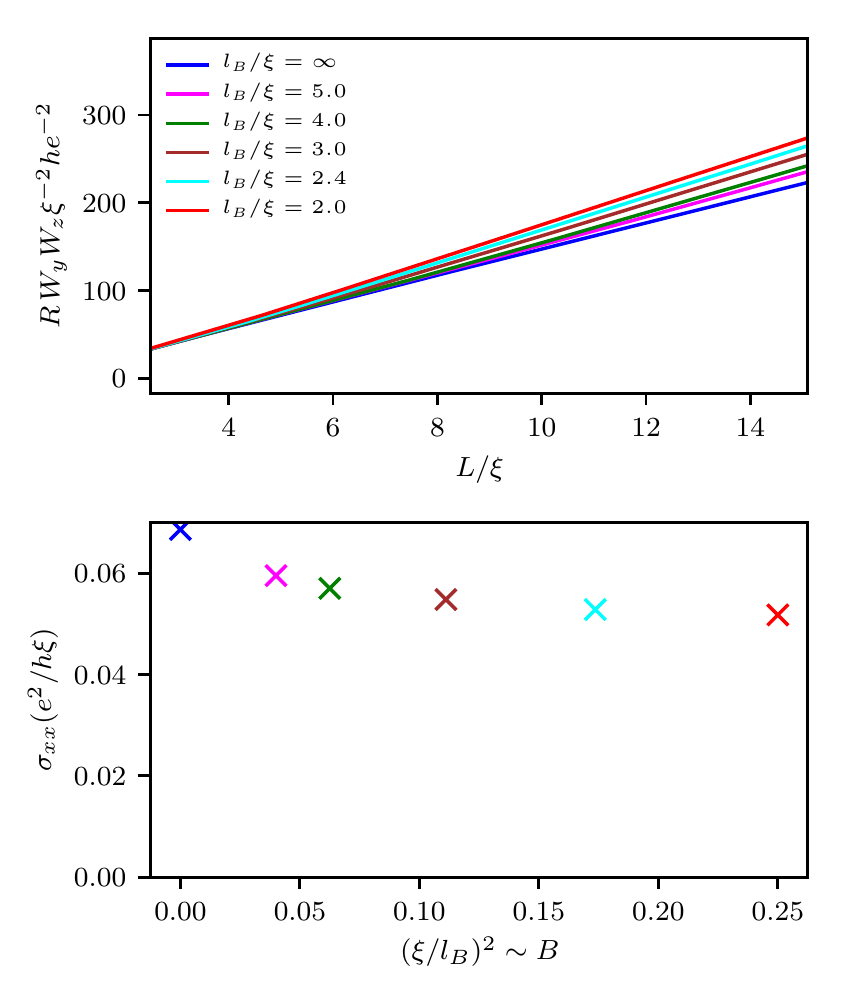}
\caption{\label{fig:strongDisorder} 
Numerical results for transversal magnetotransport in a strongly disordered Weyl node. We use $K=12$, which is well above the $B=0$ critical disorder strength. Top panel: Resistance averaged over $\sim\!1000$ disorder realizations vs.\ length. The bottom panel depicts the conductivity in the bulk limit.
}
\end{figure}

In the case of strong (supercritical) disorder, the Weyl node at $B=0$ behaves as a diffusive metal with a finite conductivity.\cite{Sbierski2014a} In Fig.~\ref{fig:strongDisorder} we show numerical results for transversal magnetotransport in the case $K=12$ that indicate a decreasing conductivity with magnetic field. Unfortunately, due to the limitations in system length, the bulk limit ($L\gg l_B$ and constant $dR/dL$) cannot be assessed for $l_B \gtrsim 5$ and the weak field scaling of $\Delta \sigma_{xx} (B)= \sigma_{xx}(B)-\sigma_{xx}(0)$ cannot be identified unambiguously. If longer system sizes become available in the future, it would be interesting to check if the predicted\cite{Klier2015} $\sim B^2$ scaling of $\Delta \sigma_{xx} (B)$ holds. For larger $B$, we observe a saturation at $\Delta \sigma_{xx} (B) / \sigma_{xx}(0) \simeq -0.25$.

\section{Conclusion \label{sec:conclusion}}
In this work, we developed a numerical approach to transversal magnetotransport in an undoped Weyl node. Our method is based on a real space formulation and is thus suitable for mesoscopic systems as well as capable of capturing the bulk limit. Building on the scattering matrix technique, we circumvent the fermion doubling theorem and faithfully describe single node physics. Our method treats both disorder and magnetic field exactly. Starting from the clean limit, the following qualitative picture emerges: The wave functions of the Landau levels are localized along the transport direction, and centered around a position determined by the crystal momentum in the direction perpendicular to both transport- and magnetic field direction. The wave functions decay exponentially in transport direction leading to a conductance exponentially decaying with system length. However, this picture is unstable under the inclusion of any finite amount of disorder, which breaks crystal momentum conservation and consequently allows for hopping between localized orbitals. Eventually, with increasing length, diffusive transport characteristics emerge.

An important aspect in any theory of transversal magnetotransport is the choice of a disorder model. The celebrated linear magnetoresistance observed in experiments is believed to be due to the presence of screened Coulomb type disorder. In this work, however, we have assumed a disorder potential with finite range correlations.

In the case of weak but finite disorder strength, our exact numerical data are in excellent agreement with results from a perturbative analytical calculation that we adapted to the disorder potential assumed.
The transversal conductivity increases linearly with weak magnetic field as long as the magnetic length is much larger than the disorder correlation lenght, $\xi \ll l_B$.
This is in agreement with previous analytical results assuming white noise disorder (i.e., $\xi=0$) in Refs.~\onlinecite{Lu2015c,Klier2015,Klier2017}.
Recently, finite range correlated disorder was also investigated analytically in Ref.~\onlinecite{Zhang2016} with the claim that it leads to decreasing transversal conductivity with increasing magnetic field if the correlation length exceeds the magnetic length, different from our analytic result in Eq.~(\ref{eq:sigma_xx}).

Further, we applied our exact numerical method to situations beyond the validity of the analytic result, namely the clean and the strongly disordered limit.

An interesting direction for future study is the straightforward generalization of our method to the aforementioned screened Coulomb disorder or to finite chemical potential.
Both modifications are relevant for experiments.
Further, in the semiclassical limit, Song et al.\cite{Song2015} proposed a guiding center picture of linear magnetoresistance that could be checked with our exact and fully quantum-mechanical approach. 

\section*{Acknowledgements}
We thank Jens H. Bardarson, Emil J. Bergholtz, Max Hering and Janina Klier for useful discussions. 
Numerical computations were done on the HPC cluster of Fachbereich Physik at FU Berlin and MPIPKS Dresden. 
F.K.K. acknowledges financial support from the Swedish research council (VR) and the Wallenberg Academy Fellows program of the Knut and Alice Wallenberg Foundation.

J.B. and F.K.K. contributed equally to this work.

\appendix
\begin{widetext}
\section{Analytical Born-Kubo calculation of transversal magnetoconductivity}

Based on Abrikosov's seminal work in Ref.~\onlinecite{Abrikosov1998} we calculate the transversal magnetoconductivity of a Weyl node assuming correlated disorder as specified in Sec.~\ref{sec:model}.
Since we want to give a comprehensive derivation, the calculation of the transversal magnetoconductivity $\sigma_{xx}$ is presented in three sections.
First, we fix conventions and introduce the clean Green function followed by a determination of the relevant disorder-induced self-energies in the Born approximation.
Finally, the Kubo formula is applied to find $\sigma_{xx}$.

\subsection{Clean Green function}

To fix the notation, we repeat the solution for the clean Hamiltonian $\mathcal{H}_0 = \hbar v \left(\mathbf{k}-\frac{e}{\hbar}\mathbf{A}\right)\cdot\boldsymbol{\sigma}$ with $\mathbf{A} = ( 0 , Bx , 0)$.
The system is placed in a box of volume $V=L^3$.
All eigenfunctions are labeled by their momentum perpendicular to the transport direction, $\mathbf{k}_\perp = (k_y,k_z)$, and the Landau-level index $m = 0,\pm1 , \pm2 ,\cdots \in\mathbb{Z}$, where negative $m$ correspond to negative energies and positive $m$ to positive energies.
For $m\neq0$, the eigenenergies and -spinors read\cite{Abrikosov1998}
\begin{align}
 \varepsilon_m (\mathbf{k}_\perp)
 &= \sgn{m} \frac{\hbar v}{l_B} \sqrt{l_B^2 k_z^2 + 2 |m|},
 & \Phi_m (\mathbf{k}_\perp) &= \frac{1}{\sqrt{2}}
 \begin{pmatrix}
 \sqrt{1+\frac{\hbar vk_{z}}{\varepsilon_{m}}} \ket{|m|, \mathbf{k}_\perp } \\
 -i \, \sgn{m} \sqrt{1-\frac{\hbar vk_{z}}{\varepsilon_{m}}} \ket{|m-1|, \mathbf{k}_\perp }
 \end{pmatrix},
\end{align}
and for $m=0$
\begin{align}
 \varepsilon_0 (\mathbf{k}_\perp)
 &= \hbar v k_z,
 & \Phi_0 (\mathbf{k}_\perp) &= 
 \begin{pmatrix}
 \ket{0, \mathbf{k}_\perp } \\
  0 \\
 \end{pmatrix},
\end{align}
with the orthonormal harmonic oscillator wave functions centered at $x_c = l_B^2 k_y$
\begin{equation}
 \phi_m (x,\mathbf{k}_\perp ) = \left\langle \left. x \right| m,\mathbf{k}_\perp \right\rangle = \frac{1}{\sqrt{l_B}} \psi_m \left( \frac{x}{l_B} - l_B k_y \right), 
 \qquad
 \int d x \phi_m (x, \mathbf{k}_\perp) \phi_{m'} (x,\mathbf{k}_\perp ) = \delta_{mm'},
\end{equation}
the Hermite functions $\psi_m (x) = H_m(x) \exp ( - x^2/2) /\sqrt{2^m m! \sqrt{\pi}}$ and the Hermite polynomials $H_m(x)$.
By counting the number of possible center coordinates $x_{c}$ within a $y$-extent of length $L$, we find that the $k_{y}$-degeneracy of each eigenenergy $\varepsilon_{m}(k_{z})$ is $g \equiv L^2/(2\pi l_{B}^{2})$.
The clean Matsubara Green function is given by
\begin{equation}
G_0 \left( i \omega_n ,m, \mathbf{k}_\perp \right) = \frac{\Phi_{m} (\mathbf{k_\perp}) \Phi_{m}^\dagger (\mathbf{k}_\perp)}{i\omega_{n}-\varepsilon_{m}(\mathbf{k}_\perp)} .
\end{equation}
The only Green functions necessary to calculate the conductivity at zero energy carry indices $m=0$ and $m=\pm 1$. Explicitly, they read
\begin{align}
G_0 \left( i \omega_n ,m=0, \mathbf{k}_\perp \right)
&= \frac{1}{i\omega_{n}-\hbar v k_z}
\begin{pmatrix}
 \ket{0,\mathbf{k}_\perp} \bra{0,\mathbf{k}_\perp} & 0 \\
 0 & 0
\end{pmatrix}, \label{eq:G0} \\
G_0 \left( i \omega_n ,m=\pm 1, \mathbf{k}_\perp \right)
&= \frac{1/2}{i\omega_{n} - \varepsilon_{\pm1}}
\begin{pmatrix}
 \left( 1 + \frac{\hbar v k_z}{\varepsilon_{\pm1}} \right) \ket{1,\mathbf{k}_\perp} \bra{1,\mathbf{k}_\perp} & \pm i \sqrt{ 1- \frac{(\hbar v k_z)^2}{\varepsilon_{\pm1}^2} }\ket{1,\mathbf{k}_\perp} \bra{0,\mathbf{k}_\perp} \\
\mp i \sqrt{ 1- \frac{(\hbar v k_z)^2}{\varepsilon_{\pm1}^2} } \ket{0,\mathbf{k}_\perp} \bra{1,\mathbf{k}_\perp} &   \left( 1 - \frac{\hbar v k_z}{\varepsilon_{\pm1}} \right)   \ket{0,\mathbf{k}_\perp} \bra{0,\mathbf{k}_\perp}
\end{pmatrix} .
\end{align}

\subsection{Disorder scattering in Born approximation}

In the Kubo calculation of the transverse magnetoconductivity $\sigma_{xx}$, which follows below, we need the imaginary part of the disorder-averaged retarded self-energy correction for $m=\pm1$ and $k_{z}=0$
\begin{equation}
 \Gamma_\pm (k_y) \equiv - \mathrm{Im} \Sigma^R_{\downarrow\downarrow} (m = \pm 1, \omega = 0, k_y , k_z = 0 ) .
\end{equation}
The Born-approximation diagram is a loop including a free propagator, which we can restrict to the $(m=0)$-Green function due to its small energy denominator.
After averaging over disorder, the diagram reads
\begin{equation}
 \Gamma_\pm (\mathbf{k}_\perp) = -\mathrm{Im} \int_{\mathbf{k'}} \braket{0,\mathbf{k}_\perp}{\mathbf{k}'} \bra{\mathbf{k}'} G_{0,\uparrow\uparrow}^{R} (0, \mathbf{k}'_\perp) \ket{\mathbf{k}'} \braket{\mathbf{k}'}{0,\mathbf{k}_\perp} \left\langle U(\mathbf{\mathbf{k}-\mathbf{k'}}) U(\mathbf{k'}-\mathbf{k}) \right\rangle_\mathrm{dis} .
 \label{eq:self_energy}
\end{equation}
The overlap between the Landau level wave function and the momentum eigenstate can be evaluated by inserting a real-space basis giving
\begin{equation}
 \Gamma_\pm (\mathbf{k}_\perp) = -\mathrm{Im} \int_{\mathbf{k}'} \frac{\int d x e^{i k_x' x} \phi_0 (x,\mathbf{k}_\perp) \phi_0 (x,\mathbf{k}_\perp') \int d x' e^{-i k_x' x'}\phi_0 (x',\mathbf{k}_\perp') \phi_0 (x',\mathbf{k}_\perp)}{i \eta - \hbar v k_z'} \left\langle U(\mathbf{k}-\mathbf{k'}) U(\mathbf{k'}-\mathbf{k}) \right\rangle_\mathrm{dis} .
\end{equation}
Inserting the disorder correlator in momentum space, i.e.,
\begin{equation}
 \left\langle U(\mathbf{q}) U(-\mathbf{q}) \right\rangle_\mathrm{dis} = K (\hbar v)^2 \xi e^{- \frac{1}{2} \xi^2 q^2},
\end{equation}
allows the evaluation of the integrals over real space and momentum giving\cite{Behrends2017}
\begin{align}
 \Gamma_\pm (\mathbf{k}_\perp)
 &= - K (\hbar v)^2\xi\, \mathrm{Im} \int_{\mathbf{k'}} \frac{e^{- \frac{1}{2} (l_B^2 + \xi^2)((k_x-k_x')^2 + (k_y-k_y')^2)}}{i \eta - \hbar v k_z'}  e^{- \frac{1}{2} \xi^2 {k_z'}^2} \\
 &= \frac{K}{4\pi} \frac{\hbar v \xi}{\xi^2 + l_B^2} e^{-\frac{1}{2}l_B^2 k_z^2} .
\end{align}
At $k_z = 0$, we finally obtain a simple expression independent of $k_{y}$ and the sign of $m=\pm1$,
\begin{equation}
\Gamma=\frac{K}{4\pi} \frac{\hbar v\xi}{l_{B}^{2}+\xi^{2}}.
\label{eq:Gamma}
\end{equation}

\subsection{Transversal magnetoconductivity from Kubo formula}

The transversal magnetoconductivity $\sigma_{xx}$ is obtained via the Kubo formula
\begin{equation}
 \sigma_{xx} = \lim\limits_{\Omega \to 0} \frac{1}{\Omega} \mathrm{Im} \Pi^R_{xx} (\Omega)
\end{equation}
with the imaginary-time current-current correlation function $\Pi_{xx} (\tau - \tau') = - \mathrm{Tr} \left[ \langle T_\tau j_x (\tau) j_x (\tau') \rangle \right]$ where $T_\tau$ denotes imaginary-time ordering and $j_x$ is the current operator $j_x = e v \,\bar{\psi} \sigma_x \psi$.
By using standards methods,\cite{BruusBook} we find
\begin{equation}
\Pi_{xx} \left( i \Omega_\ell \right) = \frac{e^{2}v^{2}}{\beta L^2} \sum_{\omega_{n}} \sum_{\mathbf{k}_\perp} \sum_{m_{1,2}=-\infty}^{\infty} \mathrm{Tr}_{\phi,\sigma} \left [\sigma_{x} G \left( i \omega_{n}+i\Omega_\ell , m_1, \mathbf{k}_\perp \right)\sigma_{x} G \left( i\omega_{n} , m_2 , \mathbf{k}_\perp \right) \right],
\label{eq:Pixx}
\end{equation}
where disorder-dressed Green functions are denoted by $G$ and the trace is over spin degrees of freedom, $\sigma$, and the Harmonic-oscillator basis, $\phi$.
In Eq.~\eqref{eq:Pixx}, we have neglected the small vertex correction, which can straightforwardly be calculated to be $\Gamma^2/(\hbar v l_B^{-1})^2$ in the lowest order.
Using rotational invariance in the $x$-$y$ plane, we write $\Pi_{xx}^R = ( \Pi_{xx}^R + \Pi_{yy}^R ) /2$ to simplify the trace to
\begin{equation}
 \frac{1}{2} \mathrm{Tr}_\sigma \left[ \sigma_x G^{(i)} \sigma_x G^{(ii)} \sigma_x + \sigma_x G^{(i)} \sigma_y G^{(ii)} \sigma_y \right] = G^{(i)}_{\uparrow\uparrow} G^{(ii)}_{\downarrow\downarrow} + G^{(i)}_{\downarrow\downarrow} G^{(ii)}_{\uparrow\uparrow}.
\end{equation}
Next, we perform the Matsubara sum using the standard procedures and expand for small $\Omega$. We find
\begin{align}
\mathrm{Im}\Pi_{xx}^{R}\left(\Omega\right) & = e^{2}v^{2}\hbar\Omega \frac{\beta}{2}\int\frac{d \omega}{2\pi} \frac{1}{L^3} \sum_{\mathbf{k}_\perp} \sum_{l}\sum_{m_{1,2}=-\infty}^{\infty} \frac{1}{\cosh^{2}\frac{\beta \omega}{2}} \nonumber \\
 & \times \bra{l,\mathbf{k}_\perp} \mathrm{Im}G_{\downarrow\downarrow}^{R} \left( \omega, m_{1},\mathbf{k}_\perp \right)\mathrm{Im} G_{\uparrow\uparrow}^{R}\left(\omega, m_{2}, \mathbf{k}_\perp \right)+(\uparrow\uparrow)\leftrightarrow(\downarrow\downarrow) \ket{l,\mathbf{k}_\perp},
\end{align}
and let $T\to 0$.
This yields
\begin{equation}
\sigma_{xx}  = \frac{\hbar e^{2}v^{2}}{\pi} \frac{1}{L^3} \sum_{\mathbf{k}_\perp}\sum_{l}\sum_{m_{1,2}=-\infty}^{\infty} \bra{l,\mathbf{k}_\perp} \mathrm{Im}G_{\downarrow\downarrow}^{R}\left(m_{1},k_{y},k_{z},0\right)\mathrm{Im}G_{\uparrow\uparrow}^{R}\left(m_{2},k_{y},k_{z},0\right)+(\uparrow\uparrow)\leftrightarrow(\downarrow\downarrow) \ket{l,\mathbf{k}_\perp}.
\end{equation}
We approximate the sums over $m_{1}$ and $m_{2}$ with the dominant terms, i.e., using the the minimal $|m_{i}|$.
This is justified in the limit of magnetic energy large compared to level width $\hbar v l_B^{-1}\gg\Gamma$ as discussed in the main text.
Due to the $2\times2$ matrix structure of $G^{R}(m=0)$ (cf.\ Eq.~\eqref{eq:G0}), the choice $m=0$ is only applicable for the $\uparrow\uparrow$-component.
Then by orthogonality of the $\left|\phi_{l}(k_{y})\right\rangle $, we only have to consider $m=\pm1$ for the remaining $\downarrow\downarrow$-Green function component:
\begin{align}
\sigma_{xx} & = \frac{2\hbar e^{2}v^{2}}{\pi} \frac{1}{L^3}\sum_{\mathbf{k}_\perp} \left( \mathrm{Im}\frac{\frac{1}{2}\left(1-\frac{l_{B}k_{z}}{\sqrt{k_{z}^{2}l_{B}^{2}+2}}\right)}{-\frac{\hbar v}{l_{B}}\sqrt{k_{z}^{2}l_{B}^{2}+2}+i\Gamma}+\mathrm{Im}\frac{\frac{1}{2}\left(1+\frac{l_{B}k_{z}}{\sqrt{k_{z}^{2}l_{B}^{2}+2}}\right)}{+\frac{\hbar v}{l_{B}}\sqrt{k_{z}^{2}l_{B}^{2}+2}+i\Gamma}\right) \mathrm{Im}\frac{1}{-\hbar vk_{z}-\Sigma_{\uparrow\uparrow}^{R}\left(0,k_{y},k_{z},0\right)}.
\end{align}
In the weak disorder limit discussed above, the last factor can be
treated as a $\delta$-function, which yields
\begin{equation}
\sigma_{xx} = \frac{e^{2}v}{2\pi}\frac{1}{L^{2}}\sum_{k_{y}}\frac{2\Gamma}{2\frac{\hbar^{2}v^{2}}{l_{B}^{2}}+\Gamma^{2}}.
\end{equation}
Along with $\sum_{k_{y}}=g=L^{2}/(2\pi l_{B}^{2})$ and the
value of $\Gamma$ determined from the Born approximation in Eq.~\eqref{eq:Gamma}, the transversal magnetoconductivity is obtained to read
\begin{equation}
\sigma_{xx}=\frac{e^{2}}{h\xi}\left( \frac{K}{8\pi^{2}}\times\frac{1}{1+l_{B}^{2}/\xi^{2}}\right) ,
\end{equation}
as quoted in the main text.
\end{widetext}

\bibliography{library}

\end{document}